\documentclass[onecolumn,amsmath,amssymb,floatfix,nofootinbib]{revtex4}

\usepackage{graphicx}
\usepackage{dcolumn}
\usepackage{bm}
\usepackage{color}

\newcommand{\be}{\begin{eqnarray}}
\newcommand{\non}{\nonumber \\}
\newcommand{\ee}{\end{eqnarray}}

\begin{document}

\title{An optimal estimator for resonance bispectra in the CMB}
\author{Moritz M\"{u}nchmeyer$^{1,2}$}
\author{P.~Daniel Meerburg$^{3,4}$}
\author{Benjamin D.~Wandelt$^{1,2,5,6}$}

\affiliation{$^1$Sorbonne Universit\'{e}s, UPMC Univ Paris 06, UMR7095}
\affiliation{$^2$CNRS, UMR7095, Institut d'Astrophysique de Paris, F-75014, Paris, France}
\affiliation{$^3$CITA, University of Toronto, 60 St. George Street, Toronto, Canada}
\affiliation{$^4$Department of Astrophysical Sciences, Princeton University, Princeton, NJ 08540 USA.}
\affiliation{$^5$Lagrange Institute (ILP) 98 bis, boulevard Arago 75014 Paris France}
\affiliation{$^6$Departments of Physics and Astronomy, University of Illinois at Urbana-Champaign, Urbana, IL 61801, USA}

\begin{abstract}
We propose an (optimal) estimator for a CMB bispectrum containing logarithmically spaced oscillations. There is tremendous theoretical interest in such bispectra, and they are predicted by a plethora of models, including axion monodromy models of inflation and initial state modifications. The number of resolved logarithmical oscillations in the bispectrum is limited due to the discrete resolution of the multipole bispectrum. We derive a simple relation between the maximum number of resolved oscillations and the frequency. We investigate several ways to factorize the primordial bispectrum, and conclude that a one dimensional expansion in the sum of the momenta $\sum k_i = k_t$ is the most efficient and flexible approach.   We compare the expansion to the exact result in multipole space and show for $\omega_{\rm eff}=100$ that $\mathcal{O}(10^3)$ modes are sufficient for an accurate reconstruction. We compute the expected $\sigma_{f_{\rm NL}}$ and find that within an effective field theory (EFT)  the overall signal to noise scales as $S/N\propto \omega^{3/2}$. Using only the temperature data we find  $S/N\sim\mathcal{O}(1-10^2)$ for the frequency domain set by the EFT. 
\end{abstract}

\maketitle

\section{Introduction}

In the last decade we have come to establish an increasingly clear picture of the early Universe through precise cosmological measurements \cite{WMAPfinal2012,PlanckCosmoPars2013}. We now know that the fluctuations produced in the early Universe are scalar-like, adiabatic, almost scale free, Gaussian and at the order of $10^{-5}$ in amplitude \cite{PlanckInflation2013,PlanckNGs2013}. Although the model that produced these fluctuations is still under debate, one can consider these stringent conditions on the type of fluctuations any model of the early Universe has to produce. This is a tremendous achievement; it is astonishing we can make such precise measurements of an event that occurred billions of years ago. Current and future experiments will have the important task to mine for new observables that can probe beyond the LCDM model; searching for traces of tensor fluctuations; non-adiabatic contributions, deviations from scale invariance, deviations from Gaussianity, and deviations from a black body through spectral distortions. All these observables can be considered extra degrees of freedom (although in most models (some of) these are correlated) which allow us to further refine our understanding of the early Universe. 

Characteristic deviations from scale invariance are a generic prediction of a broad class of models \cite{MonodromySilverstein2008,FeaturesFromHeavyPhysicsAna2011,2fieldFeatures2013,TransPlanckianMartin2001,GreeneEtAl2005}, that also produce potentially measurable levels of non-Gaussianity\cite{ResNGsFlaugerPajer,NonBDBispectrum2009,NonBDBispectrum2010,NonBDBispectrum2010b,NGFeaturesChen2007,FeaturesFromSoundSpeedAna2014,nonBDbospectrumPAgullo2011}. The details of the models differ, but we can broadly distinguish two classes\footnote{In principle one could think of other possibilities. The models presented here appear naturally.}: one in which the inflaton action is modified at a fixed time, and one in which the action is modified by a fixed (frequency) scale in physical time. 

When inflation is modified at a fixed time, the resulting feature has linearly spaced oscillations. The linearity of the oscillation is accompanied by a damping, because at the level of the action the modification must be sudden, or at the very least localized. All modes will be in a different stage of their evolution and hence will be affected differently. For example, if one modifies the equation of state (i.e. the slow roll parameter) at some time during inflation, this leads to linear decaying features in the correlation function \cite{StepTheoryJoyEtAl2007,EOSchangeBattefeldEtAl2010}. Similarly, in boundary effective field theory, there is a fixed time at which the effective prescription of inflation breaks down \cite{GreeneEtAl2005}. This can lead to linear oscillations in the correlation function through a modification of the initial state at that time,  of which the amplitude decays as a function of scale (smaller scales are further away from the hyper surface). 

When inflation is excited at a fixed frequency/scale in physical time, oscillations will be logarithmically spaced \cite{UnwindingInflation2013,MonodromySilverstein2008}. This can be best understood by realizing that the a modification of the form $e^{i\omega t }$ are logarithmic in conformal time $\tau$ and convert to logarithmic features in $k$ at freeze-out. For example, in axion monodromy has a characteristic frequency (the symmetry breaking scale or axion decay constant). The effective sinusoidal potential, which explicitly depends on this scale, leads to non-decaying logarithmically spaced oscillations in the power and higher order spectra. Similarly, in the New Physics Hyper-surface (NPH) scenario \cite{2002PhRvD..66b3511D}, there is a scale at which the Bunch Davies vacuum breaks down. This again results in logarithmically spaced oscillations. These examples are for single clock inflationary models, but similar examples exist when multiple fields are present \cite{FeaturesFromHeavyPhysicsAna2011,2fieldFeatures2013}.

Consider the following very simple derivation \cite{2002PhRvD..66b3511D,GreeneEtAl2005}. In any inflationary background, we can write down the solutions that solve the e.o.m. for the inflaton $\phi$ as a linear combination of growing and decaying modes, i.e., 
\be
\phi_k = N(k)\left[u_k + \beta(k) u_k^*\right], \nonumber \\
\phi_k^* = N(k)\left[u_k ^*+ \beta(k)^* u_k\right].
\ee
The normalization is such that $|N(k)|^2 = 1/(1-|\beta(k)|^2)$. In canonical model, in order to match our solutions to Bunch Davies in the infinite past one usually picks the decaying mode as the solution, i.e. $\beta(k) =0$, where $\beta$ is referred to as a Bogolyubov parameter which allows one to rotate the initial condition into a non-Bunch Davies state (still a pure state though, see e.g. \cite{MixedStateAgullo2011} for mixed states). The power spectrum of fluctuations for the above inflaton are computed through $P_{\phi}(k) \propto |\phi \phi^*|$, i.e. 
\be
P(k) \propto \frac{1}{1-|\beta(k)|^2} \left(1+ |\beta(k)|^2+e^{2i\psi}\beta(k)^* + e^{-2i\psi}\beta(k)\right)|u_k|^2
\ee 
where we wrote the complex mode as $u_k = |u_k|e^{i\psi}$. The spectrum is computed at horizon crossing, when the comoving mode  $k = aH$, hence the phase is $k$ independent. In the assumption that the corrections are small ($\beta \ll 1$, we can write
\be
P_{\phi}(k) \simeq P_0\left(1+2 |\beta(k)| \cos(\alpha(k) + \psi)\right)
\ee
where $P_0$ corresponds to the usual spectrum without modifications. Here we wrote the complex Bogolyubov parameter as $\beta(k) = |\beta|e^{i\alpha(k)}$. This very simple derivation shows an effective oscillation to the power spectrum, induced by a resonance between the decaying and growing modes. In e.g. in NPH $\alpha \propto \Lambda/H$, where the inflaton is excited at a fixed scale $\Lambda$. During inflation $H(k) \propto 1/\log k$ and hence the resulting oscillation is logarithmically as expected. For axion monodromy, the solution the e.o.m. lead to an effective Bogolyubov coefficient with a phase $\alpha_k \propto \phi_k/f$ \cite{PSOscillationsFlauger2013,ResNGsFlaugerPajer}, with $f$ the axion decay constant while $\phi_k$ grows logarithmically in $k$. Generally speaking then, if one is able to write the solution to the e.o.m. as an effective Bogolyubov transformation (or more complex rotation) with a time or scale dependent phase, features will appear in the power spectrum through this line of argument \footnote{In principle, for an oscillating potential the growing solution by itself might have an oscillatory component. For the axion monodromy  for example, this component is subdominant. However, this is not true for higher order correlation function, as we will argue later. }. 

The argument above is more complicated for higher order correlation functions, which measure the (self) interactions of the field. The argument for a linear or logarithmically spaced oscillation holds for higher N-point function, but although again quite generally, the solutions to the e.o.m. can be written as a superposition of growing and decaying modes, the interaction Hamiltonian determines the precise scale dependence of the features as well as which mechanism causes the oscillation. For example, in axion monodromy resonance between the wavelength of the fluctuations and the background dominates compared to the resonance between the decaying and growing modes (and hence the term that goes as $\log k_t$ dominates the term that goes as $\log K_j$, with $K_j = k_t -2k_j$). Since initial state modifications and an oscillating potential are truly different physical effects, they can actually appear at the same time, and under some circumstances enhance one another as was studied in Ref.~\cite{ResonantAndNonBDChen2010}. The measured frequency and amplitude in the power spectrum and the bispectrum (and higher order spectra) will be correlated and a combined analysis of these features would give profound insight in the model governing the early Universe. 

Unfortunately the oscillating nature of the corrections to scale invariance render these difficult to detect; even the well constrained power spectrum could still contain features that have been undetected thus far \cite{PSOscillationsMartin2004,PSOscillationsHamann2007,PSoscillationsPahud2009,PSOscillations2011,PSOscillationsBenetti2011,PSfeaturesHazra2013,PSfeaturesDvorkin2011,PSOscillationsAich2013,Meerburg2014a,Meerburg2014b,PSOscillationsFlauger2013,BICEPoscillations2014,DriftingOscillations2014}\footnote{In recent work by the authors of Ref.~\cite{DriftingOscillations2014} it was pointed out that drifting of the frequency of the oscillations could result in a non-detection. A preliminary analysis of the  Planck 2013 data was performed but no significant evidence was found, however only part of parameter space was explored. }. For the bispectrum this is even more true; the signal to noise in each mode is generally small and one relies on measuring the overlap of a model template with a data vector containing all observed modes. As a consequence, applying the wrong template leads to an underestimate of the non-Gaussian signal. In fact, if the templates are completely orthogonal the measured bispectrum would be zero. In the case of an oscillating bispectrum, the overlap with other templates is typically very small.  As a result, even if current estimates suggest that the fluctuations are Gaussian, if resonance has occurred in the early Universe, large non-Gaussianities could have been produced, and, until now,  remained undetected. It must be noted for single field models of inflation within the context of an effective field theory one generally expects signal in the higher order correlation functions will be smaller than the signal in the power spectrum \cite{EFTOscillations2011}, however multi-field counter examples exist \cite{LargeBispectrumGreen2012} where the introduction of an extra degree of freedom allows the power spectrum features to be suppressed compared to the bispectrum. For non-BD vacua, no such bound exists and higher order correlation functions can have an enhanced signal \cite{InitialStateOriginalHolman2007}. Ultimately however, even for the case of a smaller signal to noise, a joint analysis of the power spectrum and higher order correlation functions will lead to a better constraint on the parameter space. 

In this paper we study the logarithmically oscillating bispectrum in detail. We examine how the projection from primordial space to multipole space restricts the range of shape parameters that can be searched for in the CMB, and on what multipole scales the oscillations are most visible. To do this, we evaluate the CMB bispectrum numerically by two different methods. We also describe an estimator for these oscillations in the CMB data. An obstacle in constraining the logarithmically oscillating bispectra in the data is the fact that these bispectra are not factorizable. We will show that an expansion in separable modes, similar to the one presented in \cite{ShellardModeExpansion2009} and \cite{BispectrumOscillations2010}, but modified to exploit the special form of the shape under consideration, is the most efficient way to factorize these type of spectra. Our approach uses a one-dimensional expansion in terms of linear oscillations, which allows to cover the frequency space with considerably less modes than a general three-dimensional expansion would need.

An optimal estimator for linear spaced oscillations has been presented in \cite{LinearOscillationsMoritz2014} and in \cite{AngularCorrelationJackson2014} the angular correlation function was considered as an alternative to the bispectrum. The developers of the mode expansion method have recently presented important work on possible issues with jointly measuring oscillations in the power and bi-spectrum \cite{ShellardBispectrumPsJoined2014}. The first constraints on oscillating bispectra using modal reconstruction (limited to very low frequencies), were presented in \cite{PlanckNGs2013}. 

The paper will be organized as follows. In section \ref{sec:logoscibispectrum} we make some preliminary observations about the logarithmically oscillating bispectrum. In section \ref{sec:logoscicalc} we calculate the CMB bispectrum exactly via a change of variables and discuss an approach to make the shape separable. In section \ref{sec:fouriermodal} we describe our estimator, and calculate the expected signal to noise and degeneracy in the CMB for the parameters under consideration. We conclude in section  \ref{sec:conclusion}.

\section{Logarithmic oscillations in the bispectrum}
\label{sec:logoscibispectrum}

The general translation and rotation invariant primordial bispectrum of the curvature potential $\Phi$ can be written as \cite{NonBDBispectrum2009}
\begin{equation}
\label{eq_primbispectrum1} 
\langle \Phi(\mathbf{k_1}) \Phi(\mathbf{k_2}) \Phi(\mathbf{k_3}) \rangle = (2\pi)^3 \delta(\mathbf{k_{1,2,3}}) B_\Phi(k_1,k_2,k_3),
\end{equation}
where the primordial bispectrum $B_\Phi$ is a function of the magnitude of the wave numbers $k_i$ and $f_{\rm NL}$ is the amplitude of the bispectrum. 

In this paper, we are primarily interested in a logarithmically oscillating primordial bispectrum of the form 
\begin{equation}
\label{eq_oscispectrum1} 
B_\Phi(k_1,k_2,k_3) = \frac{6 \Delta_\Phi^2 f_{\rm NL}}{(k_1k_2k_3)^2} \sin\left({\omega \log \frac{k_t}{k_*} + \phi}\right).
\end{equation}
where $k_t = k_1+k_2+k_3$. As explained in the introduction, oscillating spectra are a result of a resonance between the fluctuations and the background, and are usually referred to as resonance non-Gaussianity. The pivot scale $k_*$ is conveniently set to be $1$ Mpc$^{-1}$ for the remainder of this paper and wave numbers $k$ are measured in Mpc$^{-1}$. The free parameters are then the oscillation frequency $\omega$ and the phase $\phi$ and the overall amplitude $f_{\rm NL}$. 

The reduced CMB multipole bispectrum is related to the primordial one through geometric projection and evolution 
\be
\label{eq_bispectrum_lll}
b_{\ell_1 \ell_2 \ell_3} &=& \left(\frac{2}{\pi}\right)^3 \int x^2 dx \prod_i \int dk_i k_i^2 B_\Phi(k_1,k_2,k_3)^{1/3} \Delta_{\ell_i}(k_i)j_{\ell_i}(k_i x)
\ee 
where the spherical Bessel functions $j_{\ell}(kx)$ are geometrical factors and the transfer functions $\Delta_{\ell}(k)$ reflect the evolution of the modes as the enter the horizon during the late time expansion history the Universe. The integral goes over conformal distance $x$, with recombination at $x_{\rm rec} \approx 14000$ Mpc.

\begin{figure}
\resizebox{0.98\hsize}{!}{
\includegraphics{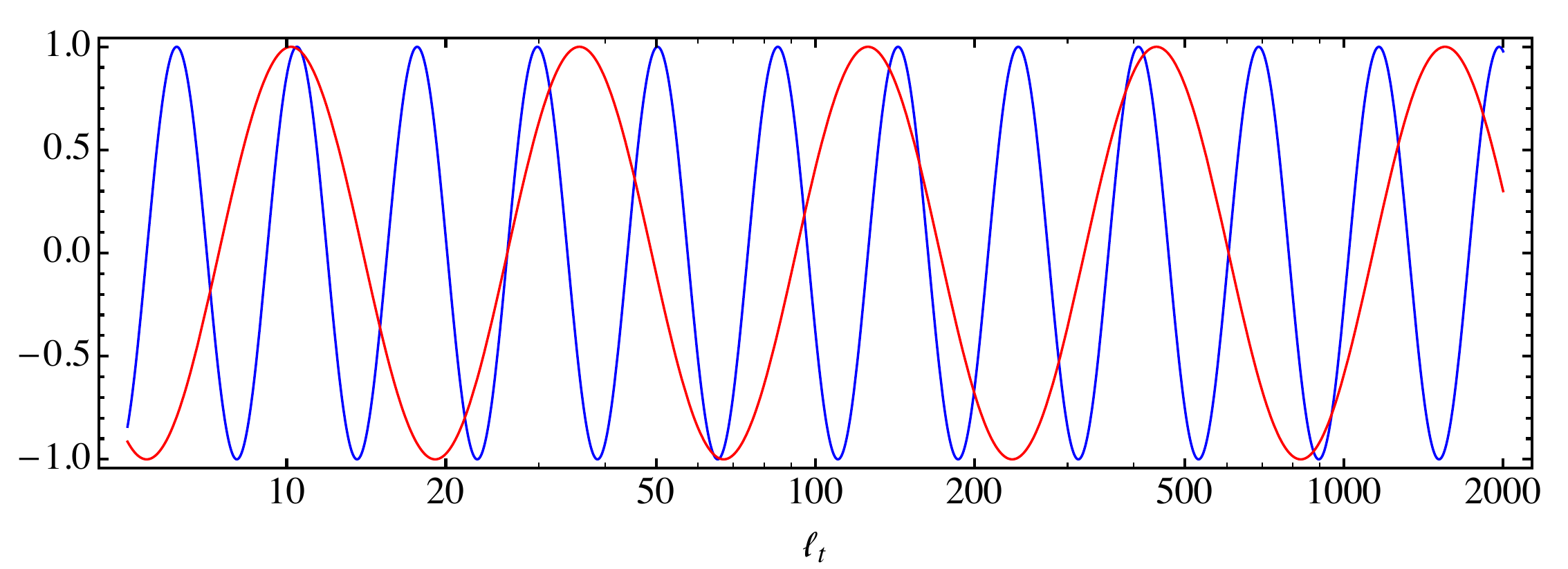}
}
\caption{Examples of oscillating primordial bispectra $\sin\left({\omega \log \frac{\ell_t}{x_{rec}} }\right)$ for $\omega=5$ (red) and $\omega=12$ (blue).}
\label{fig:logosci_examples}
\end{figure}

\begin{figure}
\resizebox{\hsize}{!}{
\includegraphics{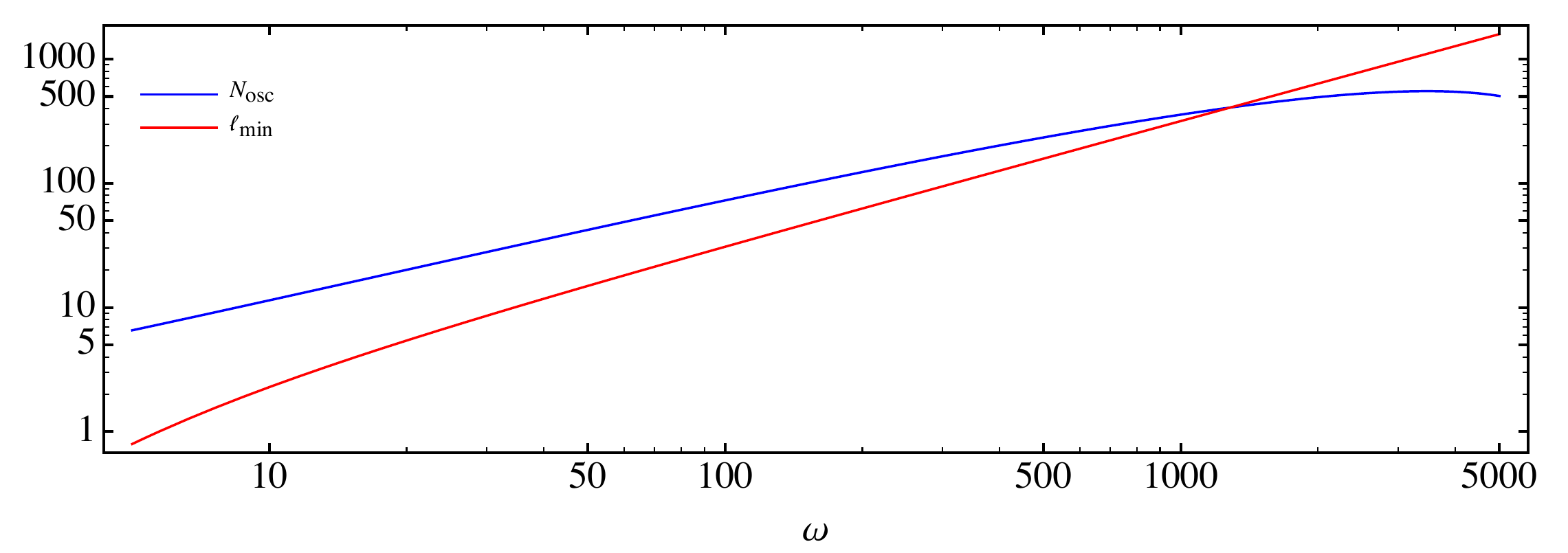}
}
\caption{Lowest multipole $\ell_{min}$ above which the oscillations can be resolved and number of oscillations $N_{\rm osc}$ resolved between $\ell_{\rm min}$ and $\ell_{\rm max}=3000$.}
\label{fig:lmin_nosc}
\end{figure}

Before calculating the CMB bispectrum for the primordial shape in eq. (\ref{eq_oscispectrum1}), we obtain a basic understanding of the possible parameter space of oscillations that can be seen in the CMB. To help intuition, we translate $k_t$ into multipoles $\ell_t$ via $\ell \simeq k x_{\rm rec}$, using the fact that the transfer functions peak strongly at recombination. We plot the primordial power spectrum (ignoring the scaling factor) as a function of $\ell_t$ in Fig.~\ref{fig:logosci_examples} for different values of $\omega$. We note that the maximum resolution that can be obtained in multipole space is $\Delta \ell = 2$. The plot thus illustrates the fact that fast oscillations, due to the logarithmic stretching, are resolved at high multipoles but not at low multipoles. More explicitly, we find that for a given frequency $\omega$ and a given minimum resolution in multipole space $\Delta \ell=2$, the oscillations can be resolved above
\be
\ell_{\rm min}(\omega,\Delta \ell) = \frac{\Delta \ell}{\left(-1 + e^{\frac{2 \pi}{\omega}}\right)}
\ee
Assuming a multipole resolution $\ell_{max}$, the total number of resolved oscillations is
\be
N_{\rm osc}(\omega,\Delta \ell) = \frac{\omega \log\left( \frac{\ell_{max}}{\Delta l} (-1 + e^\frac{2 \pi}{\omega}) \right)}{2 \pi}
\ee
These functions are shown in Fig.~\ref{fig:lmin_nosc}. We note that a maximum of $N_{\rm osc}=552$ is obtained at $\omega=3576$, for which $\ell_{\rm min}=1102$.

Some experimental constraints have already been set on logarithmic oscillations in \cite{PlanckNGs2013}. However, a large part of the parameter space has not yet been probed. 

\section{Calculation of the CMB bispectrum}
\label{sec:logoscicalc}

We now discuss how to solve the integral in eq. (\ref{eq_bispectrum_lll}). If $B_\Phi(k_1,k_2,k_3)$ is factorizable, the number of computations scales as $\ell^3$, while for an non-factorizable shape the computation scales as $\ell^{5}$. Hence, factorizability is key for the computation to be trackable. The shape of eq. (\ref{eq_oscispectrum1}) is clearly not factorable, given the factor $k_t$ within the logarithm. In the following we will discuss several possibilities to tackle this problem.

For notational simplicity we now assume that
\be
B_\Phi(k_1, k_2, k_3) &=& \frac{\sin (\omega \log k_t/k_{\star})}{k_1^2k_2^2k_3^2},
\ee
omitting the phase and the amplitude.

\subsection{Separability via an integral representation}

First, as was suggested in \cite{BispectraSmith2006}, in principle one can write any non-factoriziable shape in integrable form, of which the integrand is of the factorable form. The full computation of the projected bispectrum scales as $n*\ell^{3}$ with $n$ the number of steps in the integration. As long as $n$ is not too large, the computation is feasible. We rewrite
\be
B_\Phi(k_1, k_2, k_3) = \frac{1}{2i} \frac{e^{i \omega \log a_t}-e^{-i \omega \log a_t}}{k_1^2 k_2^2k_3^2},
\ee
with $a_t = k_t/k_*$ and realize that $e^{\pm i \omega \log a_t} = e^{\log a_t^{\pm i \omega}} = a_t^{\pm i \omega}$. An example of a factorizable integral representation is 
\be
k_t^{\pm i \omega a_t} &=& \frac{k_t}{\Gamma(1\mp i \omega)} \int_0^{\infty}t^{\mp i \omega} e^{k_t \times t}dt.
\ee

\begin{figure}
\resizebox{1.1\hsize}{!}{
\includegraphics{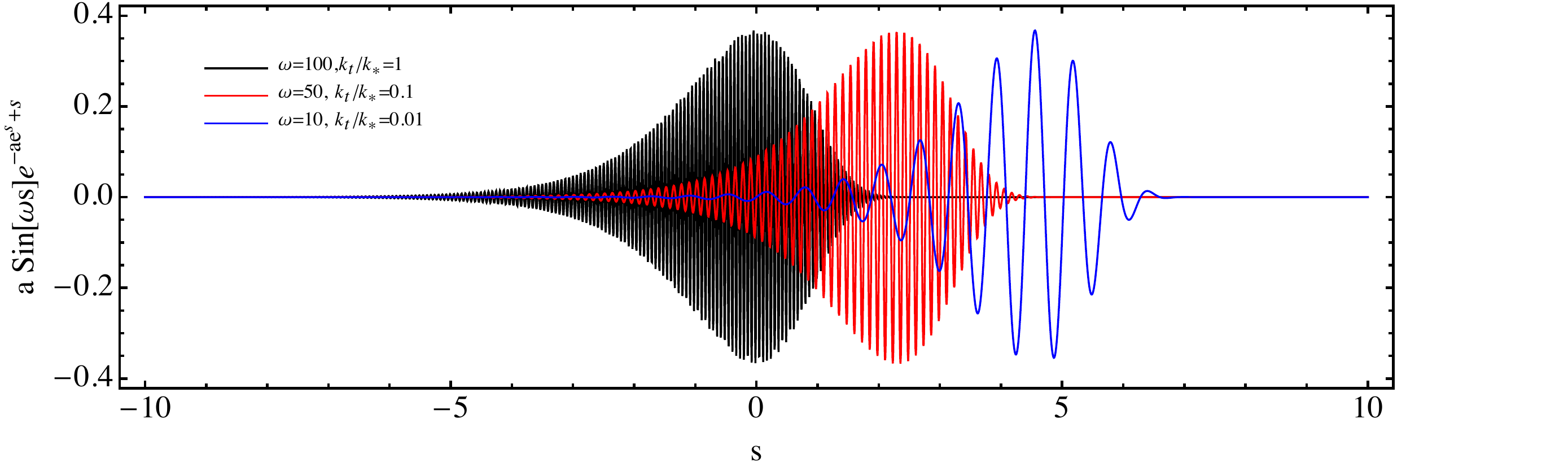}
}
\caption{The integrand of the expansion for $\sin(\omega \log a_t)$. Although this function appears integrable, numerically it is very unstable and not usable for an efficient KSW estimator, prompting us to explore other approaches. }
\label{fig:SmithZaldarriagaExpansion}
\end{figure}

We can simplify this further. The gamma functions can be written as 
\be
\Gamma(1+i\omega) &=& \Gamma(1-i\omega)^* = | \Gamma(1+i\omega)| e^{i\Theta}.
\ee
The norm of the gamma functions can be derived and we obtain
\be
 | \Gamma(1+i\omega)|^2 = \frac{\pi \omega}{ \sinh \pi \omega}.
\ee
The phase $\Theta$ can also be derived and is given by
\be
\Theta = \frac{i}{2} \log \left(-\frac{\Gamma(i\omega)}{\Gamma(-i\omega)}\right).
\ee
Note that because $\Theta$ is real, the $\log$ has to be imaginary, i.e. 
\be
\frac{\Gamma(i\omega)}{\Gamma(-i\omega)} = e^{i s},
\ee 
and $s$ has to be computed numerically.
Putting all this together, we find 
\be
\sin(\omega \log a_t) &=& \left(\frac{\pi \omega}{\sinh \pi \omega}\right)^{1/2} a_t \int_{0}^{\infty} \sin( \omega \log t - \theta) e^{-a_t t} dt,
\ee
with the phase derived from the ratio above. We can further transform this integral, first making the substitution $\Theta = \omega \log \tilde{\Theta}$, then $t = \tilde{t} \tilde{\Theta}$ and eventually $\tilde{t} = e^{s}$. We find
\be
\sin(\omega \log a_t) &=& \tilde{\Theta} \left(\frac{\pi \omega}{\sinh \pi \omega}\right)^{1/2} a_t \int_{-\infty}^{\infty} \sin(\omega s) e^{-a_t e^s +s} ds.
\ee
Thus, we have successfully transformed our logarithmically oscillating function into an infinite sum over linear oscillations, with a factorizable weighting function, $e^{-a_t e^s +s}$. We show the integrand in Fig.~\ref{fig:SmithZaldarriagaExpansion}. Although this representation is factorizable, for this method to be feasible, we require the number of steps in $s$ to be small. We find that for this representation, this not possible. Hence, although one can lower the computational cost from $\ell^5$ to $n\ell^3$, $n$ is still too large. The source of this problem lies in the power exponential, which causes nearby points to have vastly different amplitudes. We have considered special integration techniques for fast oscillating functions, but have not been able to significantly reduce the computational costs. As such, we can not apply the above parametrization to build an efficient KSW estimator. 

\subsection{Separability via a change of coordinates}

Alternatively, one can try and rewrite the integral through a coordinate transformation, under which the integral in eq. (\ref{eq_bispectrum_lll}) becomes factorable form for this shape. Such a transformation exists, and was proposed in \cite{ShellardBispectrum2006}
\be
k_1 &=& ka = k(1-\beta), \nonumber \\
k_2 &=& kb = \frac{1}{2}k(1+\alpha+\beta), \nonumber  \\
k_3 &=& kc =  \frac{1}{2}k(1-\alpha+\beta). 
\ee 
Note that $k_1 + k_2 + k_3 = k_t = (a+b+c) k = 2k$, hence $k = k_t/2$. The parameters have the following domains $0 \leq k \leq \infty$, $ 0 \leq \beta \leq 1$ and $-(1-\beta) \leq \alpha (1-\beta)$. The volume element can be computed through the determinant of the Jacobian, i.e. ${\rm Det}\;J_{ij} = k^2$, and $dk_1dk_2dk_3 = k^2 dk d \alpha d\beta$. 

For this particular shape, the integral becomes 
\be
b_{\ell_1 \ell_2 \ell_3} &=& \left(\frac{2}{\pi}\right)^3 \int d\alpha d\beta (\alpha, \beta) I^{\rm T}(\alpha, \beta) I^{\rm G} (\alpha,\beta),
\ee
with
\be
I^{\rm T} (\alpha, \beta) &=&  \int \frac{dk}{k} \Delta_{\ell_1}(ak)\Delta_{\ell_2}(bk)\Delta_{\ell_3}(ck) \cos{\omega \log 2 k}, \\
I^{\rm G} (\alpha, \beta) &=& \int x^2 dx j_{\ell_1}(a x)  j_{\ell_2}(b x)  j_{\ell_2}(c x), \\
\ee
where we used
\be
B_\Phi(k_1,k_2,k_3)  =  \frac{1}{k^6} \frac{1}{(abc)^2} \cos \omega \log 2 k.
\ee

While this form of the integral reduces the computational cost of evaluating the bispectrum, it cannot be used to construct a KSW type estimator, which avoids the costly summing over all $a_{\ell m}$ triplets. To construct such an estimator, we thus need yet another approach.  We will however use this parametrization as an exact result of the integral above which allows us to test our expansion.

\subsection{Evaluating the integral}
\label{sec:evalint}

\begin{figure}
\resizebox{1\hsize}{!}{
\includegraphics{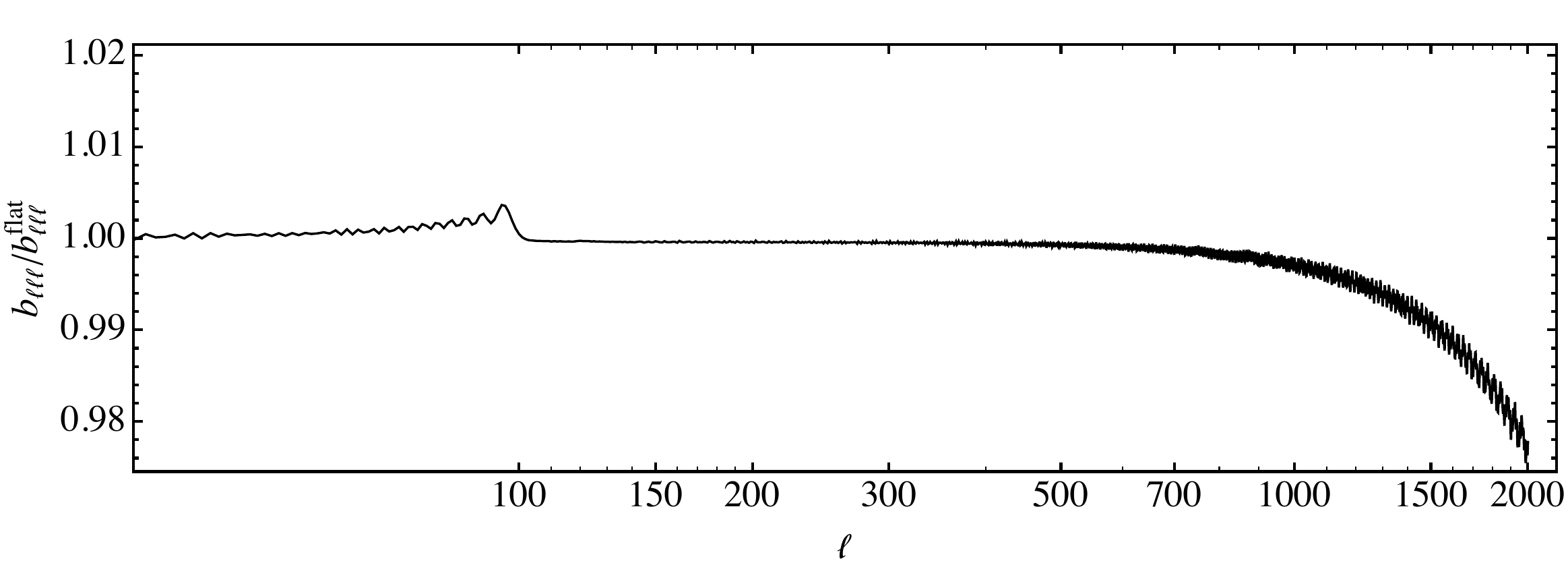}
}
\caption{The ratio between the analytical SW approximation and the numerical equivalent. As pointed out by Fergusson and Shellard, one gets a deviation at low high $\ell$ due to early truncation. For our purposes, these errors are small enough, as we only need these computations to compare with our expansion. This spectrum was generated with 1000 steps in the $\alpha-\beta$ plane. }
\label{fig:analiticVSnumeric}
\end{figure}

We use the parametrization above in order to compute the integral for the logarithmic oscillations since it drastically simplifies the computation. First, however, we need make sure this integral actually produces known results. In order to address the accuracy of our code, we use the following approximation for the flat shape\footnote{Note to be confused with the enfolded \cite{NonBDBispectrum2009}, folded or sometimes called flattened shape, which present a flattened triangle in comoving moment space, or, a shape that maximizes in the limit $k_i = k_j+k_k$, with $k_i\neq k_j\neq k_k$.} in the SW limit. 
The flat shape is given by 
\be
B_\Phi^{\rm flat}(k_{1},k_{2},k_{3})  = \frac{1}{k_1^2 k_2^2 k_3^2}.
\ee
The easiest way to compute the above integral is to perform the integrals
over $k$ first. They are finite and given by
\be
\int_{0}^{\infty}dkj_{\ell}(kx)j_{\ell}(k\Delta\eta_{*}) & = & \frac{\pi}{2(\Delta\eta_{*})^{\ell+1}}\frac{x^{\ell}}{1+2\ell}\;\mathrm{for}\; x<\Delta\eta_{*},\nonumber \\
\int_{0}^{\infty}dkj_{\ell}(kx)j_{\ell}(k\Delta\eta_{*}) & = & \frac{\pi}{2 x^{\ell+1}}\frac{\Delta\eta_{*}^{\ell}}{1+2\ell}\;\mathrm{for}\; x>\Delta\eta_{*}.\label{eq:integral1}
\ee
Consequently we can cut the integral over $x$ into two pieces, the
first one running from $\eta_{*}<x<\Delta\eta_{*}$ and the second
one running from $\Delta\eta_{*}<x<\eta_{0}$. This leads to the following
result
\be
b_{\ell_{1}\ell_{2}\ell_{3}}^{\rm flat} & = & \frac{1}{(2\ell_{1}+1)}\frac{1}{(2\ell_{2}+1)}\frac{1}{(2\ell_{3}+1)}\left\{ \frac{1}{\ell_{t}+3}\left[1-\left(\frac{\eta_{*}}{\Delta\eta_{*}}\right)^{\ell_{t}+3}\right]+\right.\left.\frac{1}{\ell_{t}}\left[1-\left(\frac{\Delta\eta_{*}}{\eta_{0}}\right)^{\ell_{t}}\right]\right\} .
\ee
Here $\ell_{t}=\ell_{1}+\ell_{2}+\ell_{3}$. Given the typical values for $\eta_{*}$
and $\eta_{0}$, $(\eta_{*}/\Delta\eta_{*})^{\ell_{t}+3}\ll1$ for all
values of $\ell_{1},\ell_{2}$ and $\ell_{3}$. 
\be
b_{\ell_{1}\ell_{2}\ell_{3}}^{\rm flat} & = &\frac{1}{(2\ell_{1}+1)}\frac{1}{(2\ell_{2}+1)}\frac{1}{(2\ell_{3}+1)}\left\{ \frac{1}{\ell_{t}+3}+\frac{1}{\ell_{t}}\left[1-\left(\frac{\Delta\eta_{*}}{\eta_{0}}\right)^{\ell_{t}}\right]\right\}.
\ee
This result is equivalent to the result found in \cite{ShellardBispectrum2006}. We use this to collaborate our code. We have modified CAMB \cite{cosmomc} by adding a module that computes this bispectrum\footnote{Send and email to daanmeerburg@gmail.com if you want a copy of the code. } using the parametrization above. As was pointed out in \cite{ShellardBispectrum2006}, because of the substructure in the tetrahedral, it requires quite a lot of sampling in the $\alpha$, $\beta$ plane to get small errors. They suggested a recursion step to speed up the calculation. It turns out that the structure of CAMB is not ideal for such an implementation, and given that we only use this for testing, we decided to run our code with high resolution instead, with 1000x1000 steps in the $\alpha-\beta$ plane. We MPI parallelized the computation and precomputing the geometrical integral for $\ell_1=\ell_2=\ell_3$ using 20 cores takes about 40 hours. Computing the alpha beta integral takes about the same. In Fig.~\ref{fig:analiticVSnumeric} we show our result for the numerical computation of the flat shape in the SW limit divided by the analytical result. As can be seen, we find good agreement up to $\ell = 1000$, when they start to deviate because of the early truncation in the Bessel functions. We use the exact result to compare our expansion used in the next section. 

\begin{figure}
\resizebox{1\hsize}{!}{
\includegraphics{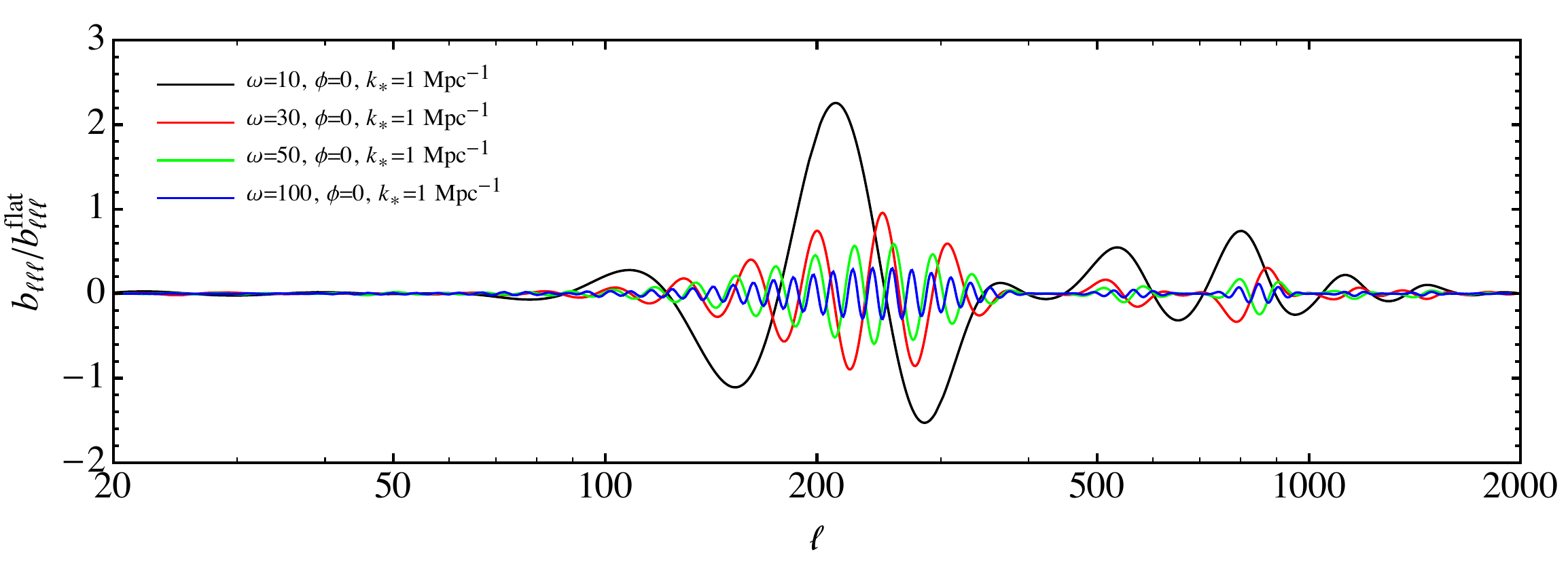}
}
\caption{The ratio between the oscillating shape for $\omega = 100$ and the flat shape. For large frequencies the ratio is always $<1$, making this shape hard to observe unless the amplitude $f_{\rm NL}$ is large. }
\label{fig:oscw100exact}
\end{figure}

\section{A Fourier expansion approach to logarithmic oscillations}
\label{sec:fouriermodal}

To create an estimator for logarithmic oscillations that is computationally tractable, we use an expansion in terms of linear oscillations. This approach is similar to the modal expansion in \cite{ShellardModeExpansion2009}, but is different in its details and more refined to this specific shape. Our approach uses the observation that the shape function in eq. (\ref{eq_primbispectrum1}) is a function of $f(k_t)$, making it effectively one dimensional in (the diagonal mode) $k_t=k_1+k_2+k_3$ coordinates and as such the expansion is more efficient. The fact that this shape is better expanded using linear oscillating modes was pointed out in \cite{BispectrumOscillations2010}. Although that setup was similar to \cite{ShellardModeExpansion2009}, including off-diagonal Fourier modes, it was already shown that only Fourier modes of $k_t$ contributed to the reconstruction.

\subsection{Fourier modal expansion of the shape function}

Any shape function of the form $S(k_t)$ can be developed in a Fourier series as
\be
S(k_t) = \sum_{n=0}^N \left(  a_n \cos{\frac{2 \pi n k_t}{\Delta k_t}} + b_n \sin{\frac{2 \pi n k_t}{\Delta k_t}} \right),
\ee
where $\Delta k$ is the supported interval and the real Fourier coefficients are
\be
a_n = \frac{2}{\Delta k_t} \int_{k_{\rm min}}^{k_{\rm max}} dk_t S(k_t) \cos{\frac{2 \pi n k_t}{\Delta k_t}}   \hspace{1cm} b_n = \frac{2}{\Delta k_t} \int_{k_{\rm min}}^{k_{\rm max}} dk_t S(k_t) \sin{\frac{2 \pi n k_t}{\Delta k_t}}, 
\ee
except for $n=0$ where the coefficients are $(\frac{a_0}{2},0)$.

The crucial property of this expansion is that the Fourier mode functions are of separable form, since $\sin(k_1+k_2+k_3)$ and $\cos(k_1+k_2+k_3)$ can be factorized in $k_1,k_2,k_3$. Here we are primarily interested in the logarithmic oscillation shape $S(k_t) = \sin\left({\omega \log(k_t) + \phi}\right)$. To make use of the Fourier expansion, the shape function must be periodic on the window $(k_{\rm min},k_{\rm max})$. We achieve this by asymptotically setting the function to zero on both sides of the $k$ window, using a generalized normal distribution of the form
\be
g(k) = e^{\left(\frac{|k-\mu|}{\alpha}\right)^\beta},
\ee
as a damping window. The primordial bispectrum expansion therefore becomes
\be
\label{eq:logosci_expansion}
B_\Phi(k_1,k_2,k_3) &=& \frac{6 \Delta_\Phi^2}{(k_1k_2k_3)^2} g(k) \sin\left({\omega \log(k_t) + \phi}\right) \\ \non
&=& 6\Big( \sum_{n=0}^N a_n \big( A_n(k_1) A_n(k_2) A_n(k_3)  - \left[ A_n(k_1) B_n(k_2) B_n(k_3)+ \textrm{2 perm.} \right] \big)  \non
 &\hspace{1cm}& + b_n \big( -B_n(k_1) B_n(k_2) B_n(k_3)  + \left[B_n(k_1) A_n(k_2) A_n(k_3) + \textrm{2 perm.}  \right] \big) \Big)
\ee
where 
\be
A_n(k) = \frac{\Delta_\Phi^{2/3}}{k^2} \cos{\frac{2 \pi n k}{\Delta k_t}}   \hspace{2cm}  B_n(k)=\frac{\Delta_\Phi^{2/3}}{k^2} \sin{\frac{2 \pi n k}{\Delta k_t}}.
\ee

\begin{figure}
\resizebox{0.7\hsize}{!}{
\includegraphics{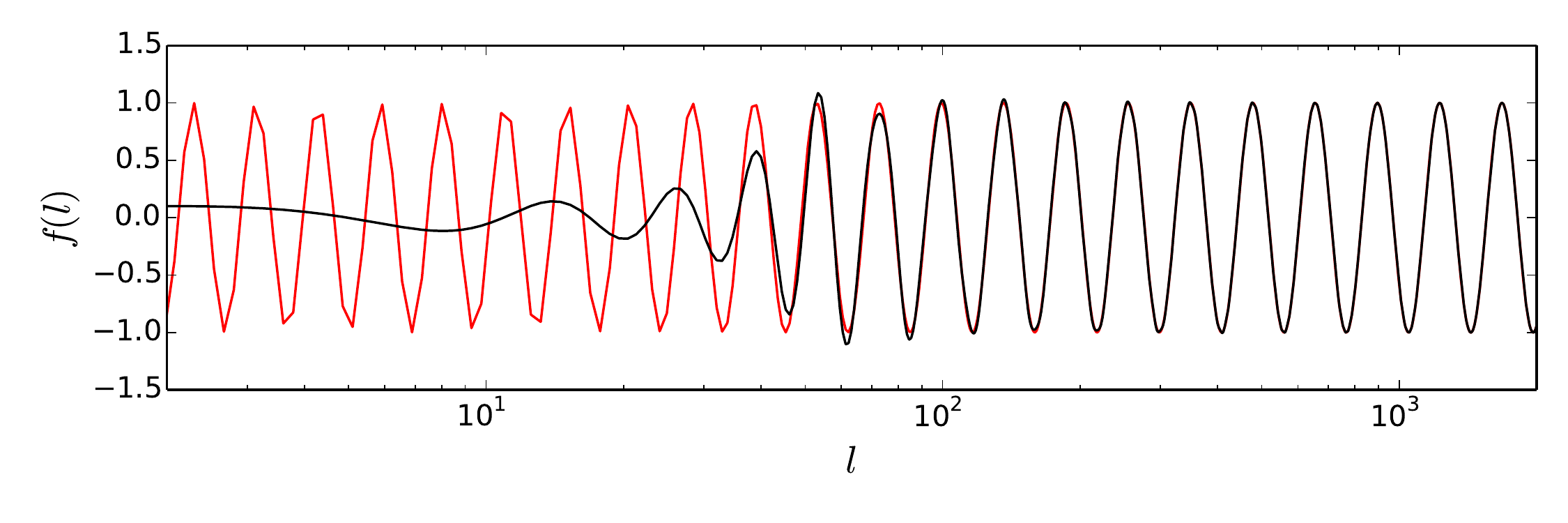}
}
\resizebox{0.7\hsize}{!}{
\includegraphics{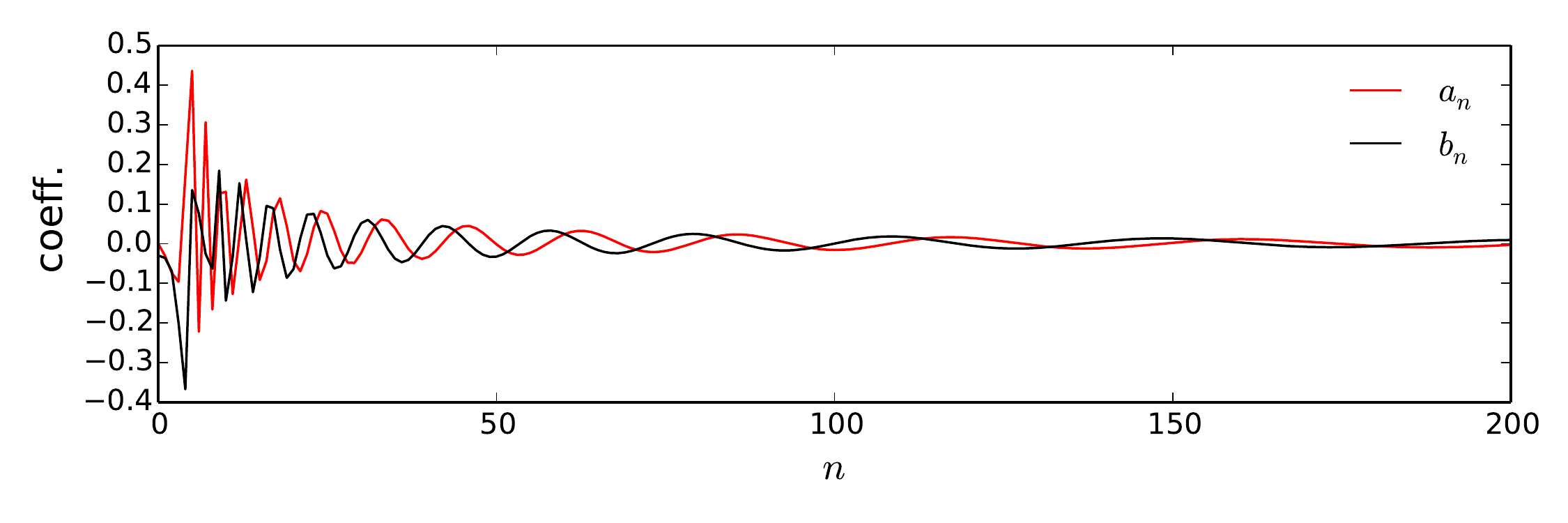}
}
\caption{Example of the shape function expansion for $\omega=10$. Top: Shape function as a function of the effective multipole scale $\ell = k x_{\rm rec}$. The red line is the original shape function, the black line is the expansion in terms of 600 Fourier modes. Bottom: Corresponding Fourier coefficients $a_n$ and $b_n$.}
\label{fig:linear_representation}
\end{figure}

The above expansion is only of practical use if the modes can be cut off at some reasonable number $N$. In the present case, we only want to resolve structures in the shape function that are larger than $\Delta \ell = 2$, as discussed in section 2. As an example, if $\ell_{\rm max}=2000$, we will therefore need about $N=1000$ modes (note that also the window function must be resolvable with this number of modes). An expansion with $600$ modes is shown in Fig.~\ref{fig:linear_representation} for $\omega=20$, using the parameters in table \ref{tab:params1}. While the range $\ell<50$ is not optimally covered, we will see below that this range does not contribute significantly to the CMB bispectrum anyway.

Let us again stress the differences of our approach to the standard modal expansion. First, this is a one dimensional expansion with expansion coefficients $a_n,b_n$, unlike the conventional modal expansion for arbitrary shapes, which has tensor coefficients $c_{ijk}$. Second, the mode functions that we use are orthogonal a priori, and consist of linear oscillations of form $\sin (\omega( k_1+k_2+k_3))$. When determining the coefficients $a_n,b_n$, we do not integrate over the usual bispectrum trapezoid in $k_1,k_2,k_3$, but rather over the one-dimensional variable $k_t$. 

Linear oscillations are by themselves a physically well-motivated shape. When searching for such linear oscillations, one has to scan the frequency space and thus has to calculate a large number of modes. It is then practical to combine a search for linear modes with a search for logarithmic modes using our approach, given that these modes are already computed to begin with.

\subsection{Results for the CMB bispectrum using the Fourier expansion}

From the separable expansion for the primordial shape function, it is straight forward to calculate the CMB multipole bispectrum. Using Eq.~\eqref{eq_oscispectrum1} and Eq.~\eqref{eq:logosci_expansion} one obtains
\be
\label{eq_bispectrum_lll_modal}
b_{\ell_1 \ell_2 \ell_3} &=& 6 \Big( \int x^2 dx  \sum_n a_n \big( A_{\ell_1}^n(x) A_{\ell_2}^n(x) A_{\ell_3}^n(x)  - \left[ A_{\ell_1}^n(x) B_{\ell_2}^n(x) B_{\ell_3}^n(x) + \textrm{2 perm.} \right] \big) \non 
&\hspace{2cm}& + b_n \big( -B_{\ell_1}^n(x) B_{\ell_2}^n(x) B_{\ell_3}^n(x)  +\left[ B_{\ell_1}^n(x) A_{\ell_2}^n(x) A_{\ell_3}^n(x) + \textrm{2 perm.} \right] \big)   \Big)
\ee 
where we have defined the functions 
\begin{align}
\label{eq_reducedbispectrum2} 
A^n_{\ell}(x) &= \frac{2}{\pi} \int dk k^2 A_n(k) j_{\ell}(kx) \Delta_{\ell}(k)\nonumber\\
B^n_{\ell}(x) &= \frac{2}{\pi} \int dk k^2 B_n(k) j_{\ell}(kx) \Delta_{\ell}(k)\nonumber\\
\end{align}
Here $j_{\ell}$ are spherical Bessel functions and $\Delta_{\ell}$ are the CMB radiation transfer functions that we evaluate numerically using CAMB \cite{cosmomc}. When evaluating the integral numerically, it must be ensured that the sampling is sufficient to resolve both the oscillations in the shape function as well as the oscillations in the Bessel functions. 

\begin{table}
    \begin{tabular}{|l|l|}
         \hline
             multipole $\ell_{\rm max}$  & 2000   \\ \hline
             primordial $k_{\rm min}$ & $10^{-5}$ \\ \hline
             primordial $k_{\rm max}$ & 0.5 \\ \hline
             modes (each sine and cosine) & 600 \\ \hline
             damping $\alpha$ & 0.12475 \\ \hline
             damping $\beta$ & 1000 \\
         \hline
    \end{tabular}
\caption{Parameters for the Fourier expansion used in this analysis.}
\label{tab:params1}
\end{table}

\begin{figure}
\resizebox{0.8\hsize}{!}{
\includegraphics{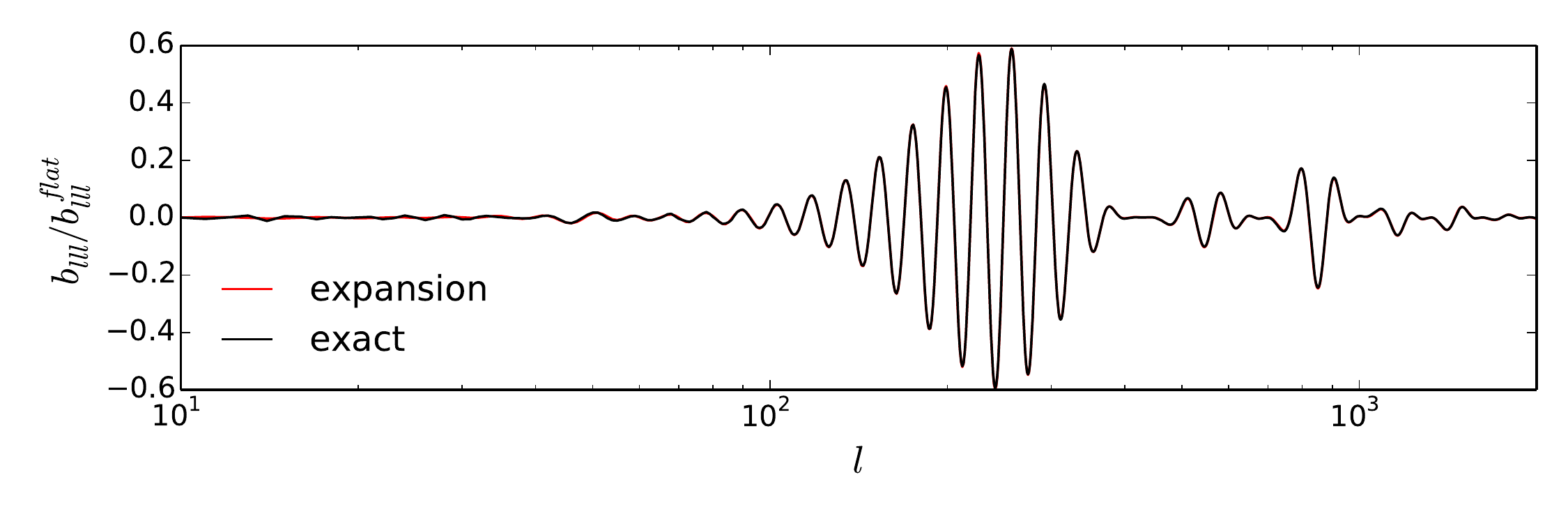}
}
\resizebox{0.8\hsize}{!}{
\includegraphics{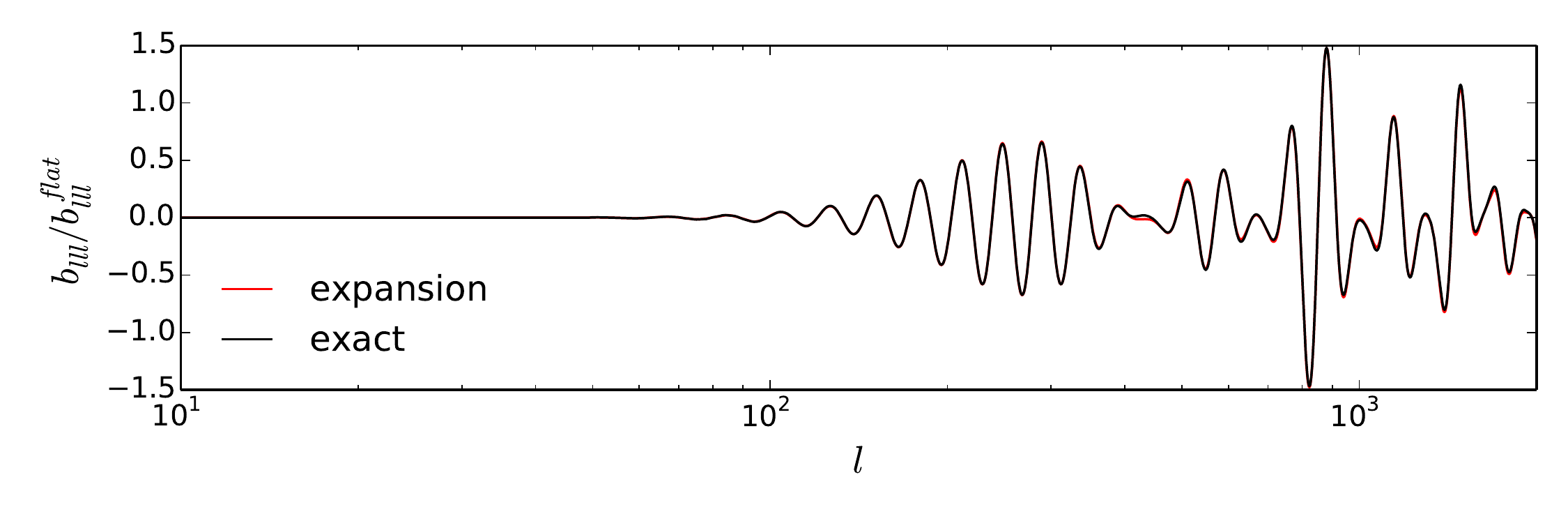}
}
\caption{Comparison between the modal expansion result and the exact result. Top: $\ell=\ell_1=\ell_2=\ell_3$. Bottom: $\ell_1=100, ~\ell=\ell_2=\ell_3$ with $\omega = 50$. Other frequencies work with comparable precision.}
\label{fig:comparison}
\end{figure}

We compare the bispectrum obtained with the Fourier expansion method to the exact results from section \ref{sec:evalint}. The parameters we used for this example are summarized in table \ref{tab:params1}. Fig.~ \ref{fig:comparison} shows the comparison for $\omega=50$. We plot again the ratio of the bispectrum to the flat bispectrum, for two different axes in multipole space, $\ell=\ell_1=\ell_2=\ell_3$ and $\ell_1=100, ~\ell=\ell_2=\ell_3$ . We find that the Fourier expansion is in excellent agreement with the exact result. We did not attempt to compare the full bispectra at all multipole combinations, since the exact numerical calculation is extremely costly. We can therefore not measure the error in the expanded bispectrum. However the one-dimensional results suggests that the expansion works as expected. 

\subsection{The KSW estimator for logarithmic oscillations}

Using the Fourier expansion in separable modes, we can now write down the KSW estimator for the CMB maps. As in the case of linear models, we can estimate amplitude and the phase explicitly by separating the bispectrum in its sine and cosine components as
\begin{equation}
\label{eq_oscispectrum3} 
B^{\mathrm{feat}}_\Phi(k_1,k_2,k_3) = \frac{6 \Delta_\Phi^2}{(k_1k_2k_3)^2} \left[ f_1 \sin\left(\ln (k_1+k_2+k_3)\right)  + f_2 \cos\left(\ln (k_1+k_2+k_3)\right)\right],
\end{equation}
where $f_{\rm NL} = \sqrt{f_1^2 + f_2^2}$ and $\Phi = \arctan{(\frac{f_2}{f_1})}$. The estimator for the two components is then given by
\begin{equation}
\label{eq_kswphase1} 
f_i= \sum_j (F^{-1})_{ij} S_j,
\end{equation}

The estimators $S_i$ are given by $S_i=S^{\rm cub}_i+S_i^{\rm lin}$ where
\begin{align}
\label{eq_ksw2} 
S^{\rm cub}_1 &=  \sum_n a_n\int r^2 dr \int d\Omega  \left[ -A^3_n(r,\hat{n}) +\left(3 A_n(r,\hat{n}) B_n(r,\hat{n}) B_n(r,\hat{n})\right) \right], \nonumber\\
S^{\rm cub}_2 &=  \sum_n b_n\int r^2 dr \int d\Omega  \left[ B^3_n(r,\hat{n}) -\left(3 A_n(r,\hat{n}) A_n(r,\hat{n}) B_n(r,\hat{n})\right) \right],
\end{align}
and
\begin{align}
\label{eq_ksw3} 
S_1^{\rm lin} =  -3 \sum_n a_n&\int r^2 dr \int  d\Omega  \bigl[- A_n(r,\hat{n}) \left< A^2_n(r,\hat{n}) \right> , \nonumber\\
&+ A_n(r,\hat{n}) \left< B^2_n(r,\hat{n}) \right> + 2 B_n(r,\hat{n}) \left< A_n(r,\hat{n}) B_n(r,\hat{n}) \right> \bigr], \nonumber\\
S_2^{\rm lin} = -3 \sum_n b_n &\int r^2 dr \int d\Omega  \bigl[ B_n(r,\hat{n}) \left< B^2_n(r,\hat{n}) \right> \nonumber\\ 
&- B_n(r,\hat{n}) \left< A^2_n(r,\hat{n}) \right> - 2 A_n(r,\hat{n}) \left< A_n(r,\hat{n}) B_n(r,\hat{n}) \right>\bigr],
\end{align}
where the expectation values have to be evaluated by Monte Carlo averaging over Gaussian map realisations generated with the same beam, mask and noise properties as expected in the data. The KSW filtered maps are given by 
\begin{align}
\label{eq_ksw1} 
A_n(r, \hat{n}) =& \sum_{\ell m} (C^{-1}a)_{\ell m}A^n_{\ell}(r) Y_{\ell m}(\hat{n}),\nonumber\\
B_n(r, \hat{n}) =& \sum_{\ell m} (C^{-1}a)_{\ell m}B^n_{\ell}(r) Y_{\ell m}(\hat{n}).
\end{align}
The Fisher matrix that normalizes the estimator is given by
\begin{align}
\label{eq_fisher2} 
F_{ij} =~ \frac{1}{6} \sum_{\ell m} &\sum_{\ell'm'}  G^{m_1m_2m_3}_{\ell_1\ell_2\ell_3}  b^i_{\ell_1\ell_2\ell_3} (C^{-1})_{\ell_1m_2,\ell'_1m'_1} (C^{-1})_{\ell_2m_2,\ell'_2m'_2} \nonumber\\
&\times(C^{-1})_{\ell_3m_3,\ell'_3m'_3}  b^j_{\ell'_1\ell'_2\ell'_3} G^{m'_1m'_2m'_3}_{\ell'_1\ell'_2\ell'_3}.
\end{align}
We compute the Fisher matrix in the next section.

The third parameter that has to be determined, besides the amplitude and phase of the oscillation, is the frequency. This parameter hast to be scanned over. The modal expansion allows to do this with a minimum of additional computational effort, since only the factors $c_n$ depend on these parameters. In the next section we study degeneracy of frequency and phase, and the sensitivity with which they can be obtained.

\subsection{Precision forecast an parameter degeneracies}

To obtain the normalization of the KSW estimator, and equivalently to determine the precision with which oscillation parameters can be obtained, one has to evaluate the Fisher matrix. For simplicity we work in the approximation of a $\ell$-diagonal $m$-independent covariance matrix $C_{\ell} = C^{CMB}_{\ell} + N_{\ell}$, where $N_{\ell}$ is the diagonal noise power spectrum of the experiment. Neglecting the off-diagonal covariance matrix elements gives almost optimal $f_{NL}$ estimates, even in the conditions of the Planck experiment. The Fisher matrix element between two logarithmic oscillation bispectra $B,B'$ of different frequency or phase is then 
\begin{align}
\label{eq_fisher3} 
F_{BB'} =~ \frac{1}{6} \sum_{l}  G^{m_1m_2m_3}_{\ell_1\ell_2\ell_3}  \frac{ (\sum_n a_n b^n_{\ell_1\ell_2\ell_3}) (\sum_m a_m b^m_{\ell_1\ell_2\ell_3}) }{C_{\ell_1}C_{\ell_2}C_{\ell_3}},
\end{align}
where we schematically wrote the mode bispectra with indices $n,m$, including both the sine and cosine contribution. The diagonal elements $(B,B)$ gives the normalization of the estimator for a particular logarithmic bispectrum $B$. The corresponding amplitude estimator $\hat{f}_{B}$ has a variance of $\sigma_f = \frac{1}{\sqrt{F_{BB}}}$. When evaluating the Fisher matrix for a large number of frequencies and phases, one would like to calculate the Fisher matrix of the mode bispectra $\mathbf{F^{mode}_{nm}}$ so that one can quickly evaluate the Fisher matrix for any mode sum by
\begin{align}
\label{eq_fisher4} 
F_{BB'} = \mathbf{a_n F^{mode}_{nm} a_m}.
\end{align}
The mode Fisher matrix for our 600 sine and cosine modes has $0.5\times1200\times1200=720.000$ elements. First pre-calculating all mode bispectra takes about 500 hours on 12 cores, and 1 tera byte of hard disc storage space. Calculating a single Fisher matrix entry between two mode bispectra takes of order of a minute. That would result in about 12000 CPU hours to calculate the mode Fisher matrix. One could reduce this by interpolation and approximations. 

\begin{figure}
\resizebox{0.5\hsize}{!}{
\includegraphics{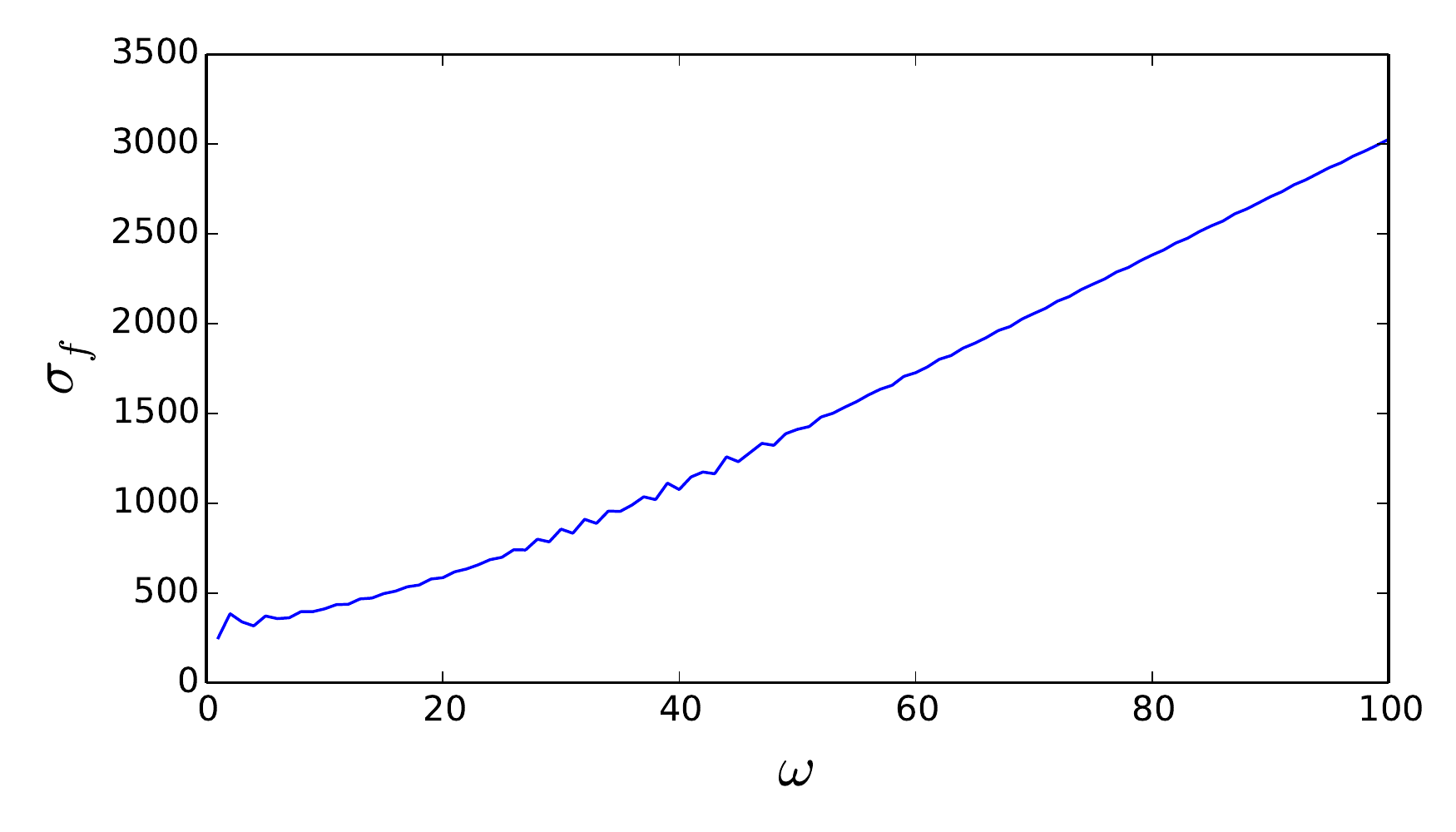}
}
\caption{Fisher forecast for the precision on the amplitude $f_{\rm NL}$ on the shape in eq. (\ref{eq_oscispectrum1}), assuming an optimal CMB experiment with $\ell_{\rm max}=2000$. We see that approximately $\sigma_{\omega} \propto \omega$.}
\label{fig:fisher1}
\end{figure}

For this paper we chose a computationally less challenging approach. We first sum the modes to calculate each $100$ logarithmic bispectra from $\omega=1$ to $\omega=100$ for sine and cosine phase. Then we calculate the Fisher matrix directly for these logarithmic bispectra, resulting in only $20.000$ Fisher matrix elements. The normalization we chose is according to eq. (\ref{eq_oscispectrum1}), where $\Delta_\Phi^2 = 9.04\times10^{-16}$. We assume full sky coverage, no noise, and $\ell_{max}=2000$. For these parameters, the sensitivity on the amplitude $f_{\rm NL}$ at a given frequency $\omega$ is shown in Fig.~\ref{fig:fisher1}, assuming for simplicity $\phi=0$. One can compare this to $\sigma_{f_{\rm NL}^{\rm flat}} \simeq 230$ for the same ideal experiment, using the same normalization (i.e. setting $\omega=0$, $\phi=\pi/2$ in the oscillating shape). Although there is some resonance between the transfer functions and the primordial bispectrum shape, the signal to noise is always smaller than the flat shape. 

\begin{figure}
\resizebox{1.0\hsize}{!}{
\includegraphics{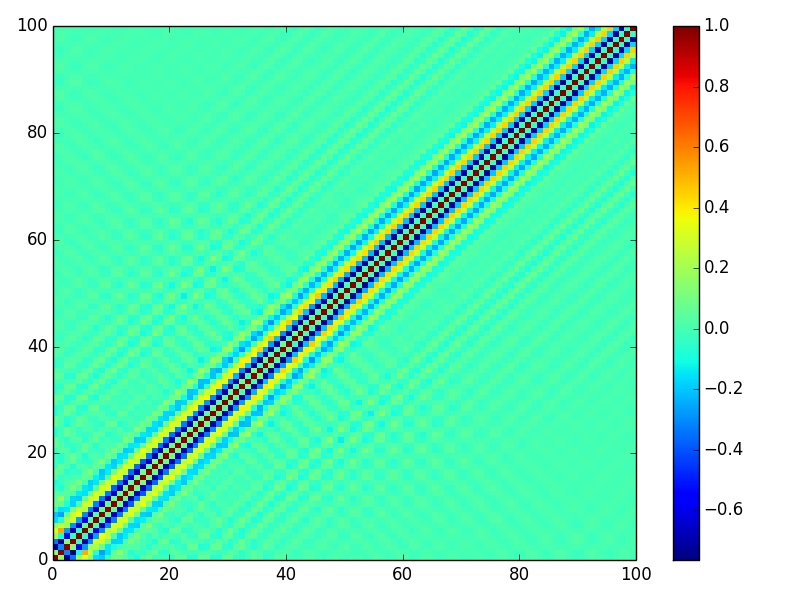}
\includegraphics{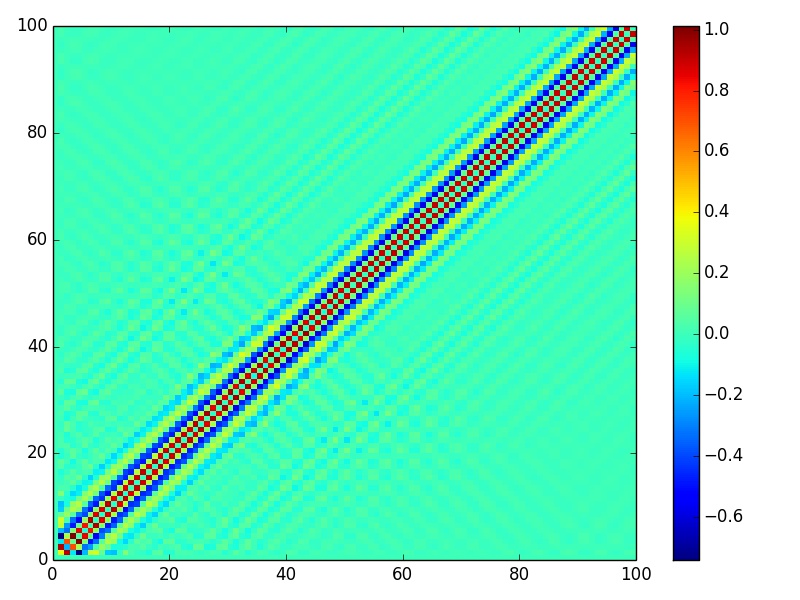}
}
\caption{Correlation matrix of logarithmic oscillations. Left: Sine to sine correlation. The frequency resolution is about $\Delta \omega = 1$. Right: Sine to cosine correlation. A strong degeneracy between neighboring frequencies and their phase is observed.}
\label{fig:fisher2}
\end{figure}

To assess how precise the frequency $\omega$ is resolved in the CMB, we plot the correlation matrix between the amplitudes $(f_i,f_j)$ of frequencies $(\omega_i,\omega_j)$, given by ${\rm corr}(f_i,f_j) = \frac{F_{ij}}{\sqrt{F_{ii} F_{jj}} }$. In Fig.~\ref{fig:fisher2} (left) this is shown for sine to sine and in Fig.~\ref{fig:fisher2} (right) for sine to cosine. The plots show that the frequency resolution is about $\Delta \omega = 1$, independent of the frequency. Further, there is a strong degeneracy between frequency and phase, as seen in the sine to cosine plot. This suggests that it is sufficient to scan for example for the sine component of the oscillation in a data set.

\subsection{Comparison to model predictions}

\begin{figure}
\resizebox{0.55\hsize}{!}{
\includegraphics{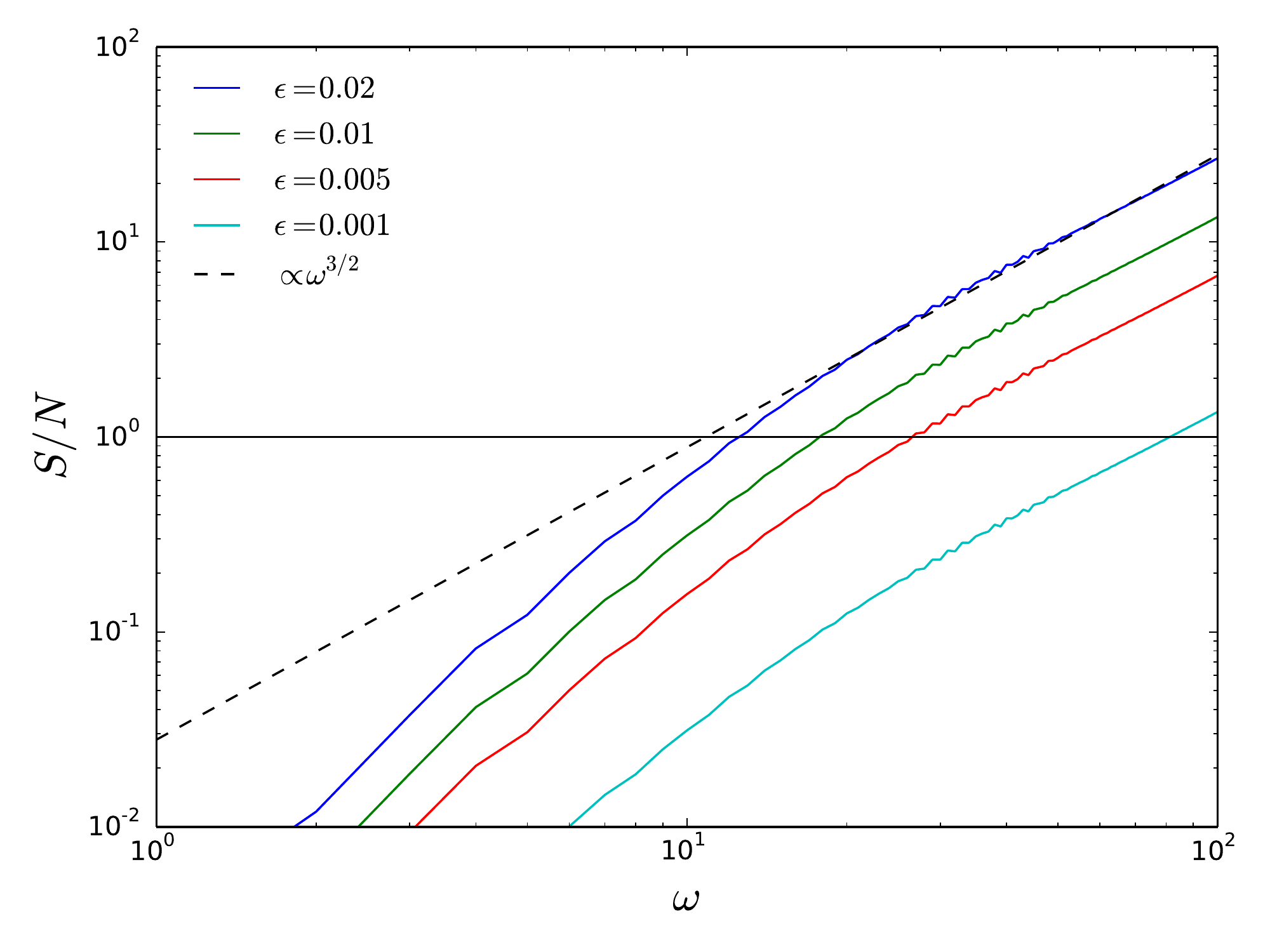}
}
\caption{Signal-to-noise ratio $S/N = f_{\rm NL} \sqrt{F_{BB}}$ for different oscillation parameters $\epsilon_{\rm osc}$ as a function of the oscillation frequency $\omega$, based on the study in \cite{EFTOscillations2011}. The dashed black line indicates the asymptotic scaling behaviour. Frequencies of order $\mathcal{O}(10^1)$ are observable, depending on $\epsilon_{\rm osc}$.}
\label{fig:signal_to_noise}
\end{figure}

We compare our result for the sensitivity on $f_{\rm NL}$ with the parameters favored by a recent study of resonant non-Gaussianities in effective field theory of inflation with a broken shift symmetry \cite{EFTOscillations2011}. This setting also includes models of axion monodromy. To be concrete, we use the shape function in Eq.~(37) of \cite{EFTOscillations2011}, neglecting the $1/\omega$ suppressed cosine term. Matching the resulting shape function to our definition in eq. (\ref{eq_oscispectrum1}), we find that in our conventions
\be
f_{\rm NL}^{\rm res} & = & \frac{1}{6} \epsilon_{\rm osc} \sqrt{2\pi}\pi^4  \omega^{5/2}.
\ee
In this model the amplitude grows as $\omega^{5/2}$. On the other hand for a fixed primordial amplitude we have approximately $\sigma_{\omega} \propto \omega$ and hence the signal-to-noise ratio increases as $S/N \propto \omega^{3/2}$. In Fig.~\ref{fig:signal_to_noise} we show the signal-to-noise for various values of the oscillation parameter $\epsilon_{\rm osc}$, which is constrained to be smaller than the slow roll parameter, i.e. $\epsilon_{\rm osc} \lesssim \epsilon_{\rm sr}$. We find that for a large range of allowed parameters oscillations are observable. Note that the perturbative control on the EFT should break down around $\omega  \sim \mathcal{O}(10^2)$ \cite{EFTOscillations2011}. 

\section{Conclusions}\label{sec:conclusion}

In this paper we developed an optimal estimator for resonant non-Gaussianities after carefully considering several options to factorize the primordial spectrum. Our expansion in a basis of linear oscillations differs from the usual modal expansion \cite{ShellardModeExpansion2009} in that it is effectively one dimensional, so only relatively few modes are needed to obtain high overlap even for very high frequencies. At the same time, the existing pipeline developed to look for linear oscillations \cite{LinearOscillationsMoritz2014} can be utilized to search for log oscillations at relatively little additional computational cost. We computed the expected signal to noise of a cosmic variance limited experiment with $\ell_{\rm max} = 2000$. We find that the $S/N \propto \omega^{3/2}$, thus although the expected signal for a constant spectrum would decay as a function of frequency, for resonant non-Gaussianity the power law behaviour of the amplitude compensates for this. Unfortunately, the frequency can not be increased indefinitely in the EFT or axion monodromy (for non Bunch Davies vacua the constraint on the amplitude is determined by back reaction, see e.g. \cite{InitialStateOriginalHolman2007}) as one loses perturbative control when the frequency gets of order $\dot{\phi}^{1/2}$. We find that within these limits $S/N\sim\mathcal{O}(1-10^2)$ just by using temperature data alone. We plan to apply the proposed estimator on Planck CMB data in a forthcoming publication.  

The example bispectrum worked out here is representative of a bispectrum that comes about in models with a discrete shift symmetry. Other models that produce logarithmic oscillations and are indistinguishable from these models at the level of power spectrum, produce different shapes for the bispectrum \cite{NonBDBispectrum2009,NonBDBispectrum2010}. To cover those shapes would for example require one to expand in $K_j = k_t - k_j$. This is more challenging, since it is effectively a 2 dimensional shape, and would therefore require new modes. It would also require a separate pipeline, since one can not expand this shape in linear oscillations as a function of $k_t$ which are part of the existing pipeline. It was shown generally that these shapes are really hard to expand in modes \cite{BispectrumOscillations2010} and further investigation is needed to find the optimal expansion for these shapes. 

In very recent work \cite{DriftingOscillations2014}, it was shown that the frequency of the oscillation in axion monodromy could drift, leading to a possible non-detection in the power spectrum when a fixed frequency template is applied to the data. The bispectrum has not been computed yet, but if the drift is still only a function of $k_t$, we expect the estimator can still be applied. In case there is additional $k$ dependence off the $k_t$ direction, our estimator might still be applicable if the drifting is factorized. 

Eventually, the estimator presented here (including polarization) should be applied jointly with estimates of the power spectrum. Such an analysis is computationally challenging; a first attempt has been made to investigate the possible presence of correlated noise in both signals in Ref.~\cite{ShellardBispectrumPsJoined2014}.We plan to further explore the possibility of jointly constraining N-point correlation functions in the future. 

We have access to unprecedented CMB temperature data, with polarization data following soon, including ground based CMB experiments to further map out the details on small scales. The effort in the next decade will be to find any deviations away from single-field slow-roll, including measurable levels of non-Gaussianity. The work presented in the paper will allow us to probe a part of parameter space which has so far been unexplored and as such will contribute in this effort.

\section*{Acknowledgments} PDM would like to thank Dan Green and Thorsten Battefeld for very useful conservations on the origin of oscillations in the power spectrum and bispectrum. 
BDW acknowledges funding from an ANR Chaire d'Excellence (ANR-10-CEXC-004-01), the UPMC Chaire Internationale in Theoretical Cosmology, and NSF grants AST-0908 902 and AST-0708849. This work made in the ILP LABEX (under reference ANR-10-LABX-63) was supported by French state funds managed by the ANR within the Investissements d'Avenir programme under reference ANR-11-IDEX-0004-02. MM acknowledges funding by Centre National d'Etudes Spatiales (CNES).

\bibliography{MDB}

\end{document}